\documentclass[aps,twocolumn,prd,preprintnumbers,superscriptaddress,nofootinbib,showpacs]{revtex4-1}%

\usepackage{amsmath}
\usepackage{amssymb}
\usepackage{graphicx}
\usepackage{hyperref}
\usepackage{mathrsfs}
\usepackage{xcolor}

\newlength{\figw}
\newlength{\figmaxw}
\def\be{\begin{equation}}
\def\ee{\end{equation}}
\def\ben{\begin{eqnarray}}
\def\een{\end{eqnarray}}

\hypersetup{
    unicode=false,          
    pdftoolbar=true,        
    pdfmenubar=true,        
    pdffitwindow=false,     
    pdfstartview={FitH},    
    pdfkeywords={keyword1} {key2} {key3}, 
    pdfnewwindow=true,      
    colorlinks=true,       
    linkcolor=red,          
    citecolor=cyan,        
    filecolor=magenta,      
    urlcolor=blue,           
    linktocpage=true
}

\newcommand{\order}[1]{\mathcal{O}\!\left(#1\right)}

\newcommand{\sss}[1]{{\scriptscriptstyle{#1}}}

\newcommand{\vev}[1]{\langle #1 \rangle}
\newcommand{\deriv}[2]{#1_{\negthinspace,#2}}
\newcommand{\ef}[1]{\tilde{#1}}

\newcommand{\eV}{\mathrm{eV}}

\newcommand{\GeV}{\mathrm{GeV}}
\newcommand{\TeV}{\mathrm{TeV}}

\newcommand{\um}{\mathrm{m}}

\newcommand{\ud}{\mathrm{d}}
\newcommand{\uPl}{\mathrm{Pl}}
\newcommand{\uEW}{\mathrm{EW}}
\newcommand{\uBD}{\mathrm{BD}}

\newcommand{\uPPN}{\mathrm{PPN}}
\newcommand{\usssPl}{\sss{\uPl}}
\newcommand{\usssEW}{\sss{\uEW}}

\newcommand{\usssPPN}{\sss{\uPPN}}
\newcommand{\usssBD}{\sss{\uBD}}

\newcommand{\uinf}{\mathrm{inf}}
\newcommand{\ucrit}{\mathrm{crit}}

\newcommand{\calL}{\mathcal{L}}
\newcommand{\calN}{\mathcal{N}}

\newcommand{\mphi}{m_0}
\newcommand{\Mp}{M_\usssPl}
\newcommand{\Mpbar}{\bar{M}_\usssPl}
\newcommand{\Mew}{M_\usssEW}

\newcommand{\OmegaL}{\Omega_{_\Lambda}}
\newcommand{\Hinf}{H_\uinf}
\newcommand{\Ho}{H_0}

\newcommand{\qn}{\eta}
\newcommand{\Vinf}{V_\uinf}
\newcommand{\phicrit}{\phi_\ucrit}
\newcommand{\Ncrit}{N_\ucrit}

\newcommand{\Npbar}{\bar{N}_\usssPl}
\newcommand{\chibar}{\bar{\chi}}
\newcommand{\phibar}{\bar{\phi}}
\newcommand{\psibar}{\bar{\psi}}
\newcommand{\betappn}{\beta_{\usssPPN}}
\newcommand{\gammappn}{\gamma_{\usssPPN}}

\newcommand{\omegaBD}{\omega_{\usssBD}}

\begin{document}

\title{A Large Mass Hierarchy from a Small Non-minimal Coupling}

\author{Christophe Ringeval} \email{christophe.ringeval@uclouvain.be}
\affiliation{Cosmology, Universe and Relativity at Louvain,
  Institute of Mathematics and Physics, Louvain University, 2 Chemin
  du Cyclotron, 1348 Louvain-la-Neuve, Belgium}

\author{Teruaki Suyama} \email{suyama@phys.titech.ac.jp}
\affiliation{Department of Physics, Tokyo Institute of Technology,
	2-12-1 Ookayama, Meguro-ku, Tokyo 152-8551, Japan}

\author{Masahide Yamaguchi} \email{gucci@phys.titech.ac.jp}
\affiliation{Department of Physics, Tokyo Institute of Technology,
	2-12-1 Ookayama, Meguro-ku, Tokyo 152-8551, Japan}

\date{\today}

\begin{abstract}
We propose a simple but novel cosmological scenario where both the
Planck mass and the dark energy scale emerge from the same
super-Hubble quantum fluctuations of a non-minimally coupled
ultra-light scalar field during primordial inflation.  The current
cosmic and solar-system observations constrain the non-minimal
coupling to be small.

\end{abstract}

\maketitle

\section{Introduction}

The standard model (SM) of particle physics and general relativity
(GR) are two pillars of the current elementary theory of physics.
Apart for non-zero neutrino masses and dark matter which, under the
new particle hypothesis, require an extension beyond the SM, there are
no observations that manifestly contradict the SM and GR.  Yet, the
wide separations among the four energy scales appearing in the SM and
GR, which are the Planck scale $\Mp=(8\pi G)^{-1/2} \simeq 10^{18} \,
\GeV$, the electroweak scale $\Mew \simeq 10^2\,\GeV$, the neutrino
mass scale $m_\nu \simeq 0.1~{\rm eV}$, and the dark energy scale
$\rho_\Lambda^{1/4} \simeq 10^{-3}\, \eV$, should provide enough
motivation to search for a dynamical explanation. One possible method
is to assume that, at least some of these quantities are not
fundamental constants, but rather fields that evolved together with
the cosmological evolution~\cite{2011LRR....14....2U}.

The idea that the gravitational constant $G$ (namely, the Planck
scale) evolves in time has long been a topic of investigation, and
many different proposals have been made in the literature in various
contexts.  For instance, Dirac was the first who conjectured that $G$
could vary with the cosmic time as $G \propto t^{-1}$ based on his
large-number hypothesis~\cite{Dirac:1937ti}. Later, scalar-tensor
theories that consistently implement the variation of $G$ were
formulated by Jordan and by Brans \& Dicke~\cite{Jordan:1959eg,
  Brans:1961sx}. Similarly, in the context of dark energy,
quintessence models have been proposed to explain the apparent
smallness of the measured dark energy density, assuming a zero
cosmological constant, and/or the coincidence
problem~\cite{Ratra:1987rm, Peebles:1998qn, Copeland:2006wr}.

In Ref.~\cite{Ringeval:2010hf}, we have shown that cosmic inflation
occurring at $\TeV$ energy scales, and therefore relatively close to
$\Mew$, could provide a natural answer to the smallness of the
cosmological constant today. The mechanism advocated there relies on
the growth of super-Hubble quantum fluctuations for an ultra-light
scalar field during primordial inflation~\cite{Grishchuk:1974ny,
  Starobinsky:1980te, Mukhanov:1981xt, Vilenkin:1982wt, Linde:1983gd,
  Starobinsky:1994bd}, which manifest themselves as a universal
quantum-generated variance after inflation. A similar mechanism for
cosmological vector fields has also been presented in
Refs.~\cite{Jimenez:2008au, Jimenez:2008nm}, again predicting an
inflationary era at the $\TeV$ scale. Various other works have since
confirmed the robustness of the mechanism and proposed extensions to
scalar-tensor theories of gravity as well as to gravitational vector
fields~\cite{Jimenez:2012ak, Glavan:2015cut, Glavan:2017jye}.

In this paper, we show that cosmic inflation can \emph{simultaneously}
explain both the largeness of the Planck scale and the smallness of
the cosmological constant by the very same mechanism: super-Hubble
quantum fluctuations of a unique non-minimally coupled ultra-light
scalar field. Proposals of an emerging Planck scale have provoked
continuous theoretical constructions within scalar-tensor theories,
but only a few have been concerned with the generation of an effective
Planck mass from inflation~\cite{1989PhRvL..62..376L,
  GarciaBellido:1993wn, GarciaBellido:1994vz, GarciaBellido:1995br,
  GarciaBellido:1995kc, Susperregi:1996hs, Biswas:2005vz}. As far as
we are aware, the scenario we propose is a new way to address the dark
energy scale and the value of he Planck mass simultaneously while
providing a potential link to the physics around the electroweak
energy scale. Let us finally mention that such a scenario is
fundamentally different compared to induced gravity
theories~\cite{Sakharov:1967pk, Visser:2002ew}, in which the
Einstein-Hilbert term emerges from the quantum fluctuations of
matter fields immersed in the curved spacetime. In our case, the
Einstein-Hilbert term is already present at the classical level,
although inflation makes it become negligibly small compared to the
non-minimal coupling term.

The paper is organized as follows. In the next section we present the
main idea and basic model requirements needed for the scenario to
work, before turning to a more detailed calculation in
Sec.~\ref{sec:calc}. In Sec.~\ref{sec:obs} we enumerate the
observational constraints, inSec.~\ref{sec:disc} we discuss other
aspects of the scenario, and we conclude in Sec.~\ref{sec:conc}.

\section{Main idea}

\label{sec:idea}

The idea relies on an ultra-light scalar field $\phi$ which only
couples to gravity with a non-minimal coupling to the Ricci scalar
$R$. During an extended period of inflation, it undergoes a
significant growth and could acquire a quantum-generated
super-Planckian variance. In the next section we will explain this
process in more detail, but here we describe the model
requirements. The relevant part of the Lagrangian is given by
\begin{equation}
\calL=\frac{1}{2} \left( M^2+\xi \phi^2\right) R-\frac{1}{2}
{(\partial \phi)}^2-\frac{1}{2} \mphi^2 \phi^2,
\label{toy-model}
\end{equation}
where $\xi$ represents the strength of the non-minimal coupling. The
hypothesis that forms the basis of this paper is that the \emph{bare}
gravitational energy scale $M$ is much smaller than the measured
Planck mass $M \ll \Mp$ and could be as low as or even smaller than
the electroweak scale. Since the main result does not depend on the
concrete value of $M$, we leave $M$ unspecified aside from the
condition $M \ll \Mp$. As Eq.~\eqref{toy-model} shows, non-vanishing
and time-independent vacuum expectation values (VEVs) for
$\vev{\phi^2}$ contribute to the effective gravitational energy scale
by $\xi\vev{\phi^2}$. We therefore require that $\xi$ be positive;
otherwise, our scenario does not work.

Once $\xi \vev{\phi^2}$ settles to Planck-like values, the potential
energy of the field today can source the acceleration of the Universe
by the mechanism of Ref.~\cite{Ringeval:2010hf}. For this to happen,
it should match the cosmological constant energy scale,
\begin{equation}
\dfrac{1}{2} \mphi^2 \vev{\phi^2} \simeq 3 \Ho^2
\Mp^2 \OmegaL.
\label{eq:lambda}
\end{equation}
Moreover, $\phi$ behaves as dark energy provided it remains (quasi)
frozen in the Hubble flow and, as discussed in Sec.~\ref{sec:obs},
this implies some constraints on $\xi$.

During inflation, due to the non-minimal coupling, the effective mass
of the field is given by
\begin{equation}
m^2 = \mphi^2 -12 \xi \Hinf^2,
\label{eq:meff}
\end{equation}
where we have taken the de Sitter value for the Ricci scalar $R=12
\Hinf^2$. There are \emph{a priori} three possible regimes.

In the limit $\xi \ll \mphi^2/(12 \Hinf^2)$, the non-minimal coupling
is so small that it has essentially no effect during inflation. The
model matches the one of Ref.~\cite{Ringeval:2010hf}, and for sufficiently
long inflation one gets the de Sitter variance of a test scalar
field, $\vev{\phi^2} \to 3\Hinf^4/(8\pi^2 \mphi^2)$. Dark energy is
explained by satisfying Eq.~\eqref{eq:lambda}, namely, for inflation
occurring at the $\TeV$ scale, $\Hinf^2 = 4\pi \sqrt{\OmegaL} \Ho
\Mp$. As a result, one gets
\begin{equation}
  \dfrac{\xi \vev{\phi^2}}{\Mp^2} \to \dfrac{3 \Ho
    \sqrt{\OmegaL}}{2\pi \Mp} \, \dfrac{\xi \Hinf^2}{\mphi^2} \ll 1,
\end{equation}
and thus the super-Hubble quantum fluctuations of $\xi \phi^2$ are always
deeply sub-Planckian and the model cannot explain the measured
Planck mass.

One could then consider the massless limit of Eq.~\eqref{eq:meff},
obtained by taking quite fine-tuned values of $\xi \to \mphi^2/(12
\Hinf^2)$. Because $m^2 \to 0$ during inflation, $\vev{\phi^2} \to
3\Hinf^4/(8\pi^2 m^2)$ can become very large. Plugging these values into
Eq.~\eqref{eq:lambda}, and requiring $\xi \vev{\phi^2} \simeq \Mp^2$,
one obtains a condition for the energy scale of inflation which, after
some algebra, reads $\Hinf^2 \simeq (\OmegaL/2) \Ho^2$, and the model is
also ruled out.

The only remaining possibility is $\mphi^2 < 12 \xi \Hinf^2$ and we are
in presence of a ultra-light tachyonic field \emph{during}
inflation. Such a situation is not problematic and has been considered
as a dark energy candidate in Ref.~\cite{Glavan:2017jye}.  Indeed,
because of the bare mass of the field $\mphi^2 > 0$, the tachyonic
instability generated by the expansion of the Universe through the
non-minimal coupling is only transient.  As we detail below, such a
transient instability is actually a virtue and allows the mechanism to
generate both the Planck mass and the actual value of dark energy.

\section{Quantum generated field variance}

\label{sec:calc}

Let us now consider the limit $\mphi^2 \ll 12 \xi \Hinf^2$ to perform
a more detailed calculation of the quantum-generated variance for
$\phi$. We moreover assume that inflation lasted for a very long time
in the sense that the total number of e-folds of accelerated expansion
can be a large number. For a slowly evolving Hubble parameter $H$
during inflation, the $\phi$ field undergoes a stochastic process on
super-Hubble scales, which effectively pushes its variance to larger
amplitudes~\cite{Starobinsky:1986fx, 1989PhLB..219..240N,
  Starobinsky:1994bd, 2015EPJC...75..413V}. Then, under the
slow-motion approximation, the coarse-grained field (which we still
denote by $\phi$ here) follows the Langevin equation
\begin{equation}
  \frac{\ud \phi}{\ud N} = 4 \xi \phi + \dfrac{H}{2\pi} \qn(N),
  \label{eom-phi}
\end{equation}
where $N=\int H \ud t$ is the number of e-fold and we have used $m^2
\simeq -12 \xi H^2$. The second term on the right-hand side represents
a stochastic noise arising from the transition of the sub-Hubble modes
to the super-Hubble modes. The quantity $\qn$ is a Gaussian white
noise whose two-point correlation function is given by
\begin{equation}
  \vev{\qn(N_1)\qn(N_2)} = \delta(N_1-N_2),
\end{equation}
with $\vev{\eta(N)}=0$. The Hubble parameter $H$ is determined by the
Friedmann-Lema\^itre equation stemming from Eq.~\eqref{toy-model},
plus other terms coming from the field driving inflation. If we denote
the inflaton field by $\psi$, where $V(\psi)$ is its potential, one gets
\begin{equation}
\begin{aligned}
  H^2 & =\dfrac{V(\psi)}{3 (M^2 + \xi \phi^2) -\dfrac{1}{2}
    {\deriv{\phi}{N}}^2 - 6 \xi \phi \deriv{\phi}{N} - \dfrac{1}{2}
    \deriv{\psi}{N}^2} \\ & \simeq \dfrac{V(\psi)}{3 (M^2 + \xi
    \phi^2)}\,,
  \label{eq:FL}
\end{aligned}
\end{equation}
where a comma denotes a derivative. The second line is obtained by
assuming slow-roll and keeping only the leading term. Assuming
$V(\psi) = \Vinf$ to be almost constant during a plateau-like
inflationary era, we can solve the Langevin equation to determine the
stochastic motion of $\phi$. Since $M$ is the fundamental scale in the
present scenario, we assume $M^4 \gtrsim \Vinf$ in the following
analysis.

The dependence of $H$ on $\phi$ prevents us from solving Eq.~\eqref{eom-phi}
exactly, but the solution can be approximated in two domains. Defining
\begin{equation}
\phicrit \equiv \dfrac{M}{\sqrt{\xi}}\,,
\end{equation}
one sees that the behavior of $H$ changes at $\phi = \phicrit$. We
exploit this observation and consider the two limiting cases $\phi
\ll \phicrit$ and $\phi \gg \phicrit$ separately, and then combine them
to obtain the (approximate) final result. In order to give a
conservative estimate, we assume that $\phi$, as well its classical
value, are initially vanishing.

Let us first investigate the motion of $\phi$ for $\phi \ll \phicrit$.
During this phase, we can ignore the term $\xi \phi^2$ in the
Friedmann-Lema\^itre equation, and the Langevin equation for $\phi$
can be solved analytically. One gets
\begin{equation}
  \phi(N)=\dfrac{1}{2\pi M} \sqrt{\dfrac{\Vinf}{3}} e^{4\xi N}
  \int_0^N e^{-4\xi N'} \qn(N') \ud N'.
\end{equation}
Thus, the expectation value of $\phi^2$ is
given by
\begin{equation}
  \vev{\phi^2(N)} = \dfrac{\Vinf}{96\pi^2 M^2 \xi} \left(e^{8\xi
    N}-1\right).
  \label{phi2-1}
\end{equation}
As it should be, this solution incorporates the features of both the
stochastic motion and the tachyonic instability. For $\xi N \ll 1$,
picking up the leading term, we obtain $\langle \phi^2 \rangle \approx
\Vinf/(12\pi^2M^2) N$ and recover Brownian motion.  For $\xi N \gg 1$,
we have $\langle \phi^2 \rangle \propto e^{8\xi N}$ and its
exponential growth represents the tachyonic instability.  Let us
notice that had we started from a non-vanishing VEV for $\phi$,
Eq.~\eqref{phi2-1} would still apply but for the variance, i.e.,
$\vev{\delta \phi^2} = \vev{\phi^2} - \vev{\phi}^2$.  If we further
add the fluctuations of $\phi$ at the initial time $\delta \phi (0)$,
Eq.~\eqref{phi2-1} contains an additional term evolving as
$\vev{\delta \phi(0)}^2 e^{8 \xi N}$.  As a result, for all possible
initial conditions, a long-enough inflationary period always induces
an exponential growth of the field variance.

However, Eq.~\eqref{phi2-1} becomes invalid when $\sqrt{\langle \phi^2
  \rangle}$ reaches $\phicrit$.  In terms of the number of e-fold,
this happens at $N=\Ncrit$, where $\Ncrit$ is given by
\begin{equation}
\Ncrit = \dfrac{1}{8\xi} \ln \left( 1+\dfrac{96\pi^2 M^4}{\Vinf}
\right).
\label{eq:Ncrit}
\end{equation}
Thus, $\Ncrit =\order{\xi^{-1}}$ and becomes very large for small
$\xi$. Next, let us investigate the opposite regime, $\phi \gg \phicrit$. In
this limit, we can ignore the term $M^2$ in the Friedmann-Lema\^itre
equation and we can solve the Langevin equation analytically for
$\phi^2$. The result is given by
\begin{equation}
\begin{aligned}
  \phi^2(N) &= e^{8\xi (N-\Ncrit)} \phicrit^2 \\ &+ e^{8\xi N}
  \sqrt{\dfrac{V}{3\pi^2 \xi}} \int_{\Ncrit}^N \eta (N')e^{-8\xi N'}
  \ud N'.
\end{aligned}
\end{equation}
The second term on the right-hand side is directly sourced by the
stochastic noise $\qn(N')$ and disappears by taking the statistical
average. Hence, one obtains
\begin{equation}
  \xi \vev{\phi^2(N)} = \dfrac{M^2}{1+\dfrac{96\pi^2 M^4}{\Vinf}} \,
  e^{8\xi N}.
\end{equation}
From this equation, we can estimate the typical number of e-folds required for
the $\phi$ field to generate a large gravitational energy scale, say
$\Mpbar$, as
\begin{equation}
\Npbar = \Ncrit + \dfrac{1}{4\xi} \ln \left( \dfrac{\Mpbar}{M} \right).
\label{eq:Npbar}
\end{equation}
Thus, $\Npbar$ is also $\order{\xi^{-1}}$.  Here we have introduced
the new mass scale $\Mpbar$ instead of the usual Planck mass $\Mp$
because, as explained in the next section, the gravitational coupling
$M^2 + \xi \vev{\phi^2}$ appearing in the Lagrangian~\eqref{toy-model}
does not necessarily equal the one measured by Cavendish-like
experiments due to the existence of a fifth force. To summarize, for
all possible initial conditions of $\phi$, a Planck-like energy scale
$\Mpbar$ can be generated by $\xi \vev{\phi^2}$ provided primordial
inflation lasts for about $\order{\xi^{-1}}$ e-folds\footnote{Strictly
  speaking, the number of e-folds $\calN$ along each trajectory is a
  stochastic quantity and another possible route for deriving the
  result is to calculate its mean stochastic value
  $\vev{\calN}$~\cite{Fujita:2013cna, 2015EPJC...75..413V,
    Vennin:2016wnk, Firouzjahi:2018vet}. For the regime $\phi <
  \phicrit$, one finds
  \begin{equation}
    \vev{\calN} \simeq \dfrac{1}{8\xi} \ln \left(\dfrac{192
      \pi^2 M^4}{\Vinf}\right) + \dfrac{\gamma}{8 \xi}\,,
  \end{equation}
  which matches $\Ncrit$ up to a factor of $\order{1}$ correction.
  Here $\gamma \simeq 0.5772$ is the Euler's constant. For the regime
  $\phi>\phicrit$, one finds
  \begin{equation}
    \vev{\calN} \simeq \dfrac{1}{4\xi} \left[
      \ln\left(\dfrac{\Mpbar}{M}\right) +
      \dfrac{\Vinf}{384 \pi^2 M^4} \right],
  \end{equation}
  which matches the second term of Eq.~\eqref{eq:Npbar} up to a factor
  of $\order{1}$ correction.}.

Let us stress that the inflationary period relevant for observations
is only about $60$ e-folds before the end and we have found that the
time scale for the variation of $\phi$ is $\xi^{-1}$ (in e-folds). As
a result, and provided inflation can end (see Sec.~\ref{sec:disc}),
the variation of $\phi$ during the last $60$ e-folds of inflation is
thus negligibly small. Standard GR is perfectly recovered during the
inflationary era relevant to observations. Let us now examine the
experimental bounds on such a mechanism.

\section{Experimental bounds}

\label{sec:obs}

The existence of an ultra-light massive field $\phi$ today leaves
various observational signatures from which we can place bounds on
both $\xi$ and $\mphi$.

Although $\phi$ is not directly coupled to matter, the non-minimal
coupling of the ultra-light scalar field induces a fifth force among
bodies, in addition to the pure GR gravitational terms. This effect
can be made manifest by making a conformal
transformation~\cite{Maeda:1988ab} from the present frame with the metric
$g_{\mu \nu}$ to the Einstein frame with the metric $\ef{g}_{\mu
  \nu}$ verifying
\begin{equation}
  g_{\mu\nu} = A^2(\phi) \ef{g}_{\mu\nu},
\end{equation}
where
\begin{equation}
  A^2 \equiv \dfrac{1}{\left(1 + \xi \dfrac{\phi^2}{M^2} \right)}\,.
\end{equation}
The action can be canonically normalized from the field redefinition
$\phi \to \chi$ with~\cite{2014PDU.....5...75M}
\begin{equation}
\begin{aligned}
  e^{\chibar} & \equiv \left[\dfrac{\sqrt{1+ \xi
        \phibar^2}}{\sqrt{1+\xi(1+6\xi) \phibar^2} + \sqrt{6} \xi
      \phibar} \right]^{\sqrt{6}} \\ & \times \left[\sqrt{1+ \left(
      \xi+6\xi^2 \right) \phibar^2} + \sqrt{\xi(1+6\xi) \phibar^2}
    \right]^{\sqrt{6+\frac{1}{\xi}}},
\label{eq:fieldEF}
\end{aligned}
\end{equation}
where we have defined the dimensionless fields $\phibar \equiv \phi/M$
and $\chibar \equiv \chi/M$. The original action is transformed as
\begin{equation}
\begin{aligned}
  S & = \dfrac{M^2}{2} \int \ud^4x \sqrt{-\ef{g}} \,
  \left[\ef{R}-\left(\ef{\nabla} \chibar \right)^2 - 2
    \dfrac{W(\chi)}{M^2} \right] \\ &+
  S_\um\left[A^2(\chi)\ef{g}_{\mu \nu},\psi_\um \right],
\end{aligned}
\end{equation}
where the potential is given by
\begin{equation}
  \dfrac{W(\chi)}{M^2} = \dfrac{A^4(\chi)}{2}\mphi^2 \phibar^2.
\end{equation}
The field redefinition \eqref{eq:fieldEF} cannot be straightforwardly
inverted, but we can take the limit we are interested in, namely $\xi
\phibar^2 = \Mpbar^2/M^2 \gg 1$ and $\xi \ll 1$. We obtain
\begin{equation}
\phibar \simeq \dfrac{1}{2 \sqrt{\xi}} \, e^{\sqrt{\xi} \chibar},
\qquad A^2 \simeq \dfrac{1}{1 + \dfrac{1}{4} e^{2\sqrt{\xi}
    \chibar}}\,.
\label{eq:appfieldEF}
\end{equation}

As it should be, the coupling between $\chi$ and matter disappears in
the minimal coupling limit ($\xi \to 0$). Such a fifth force changes
the parametrized post-Newtonian (PPN) parameters compared to the
values in GR as~\cite{Damour:1992we, Hohmann:2013rba, Jarv:2014hma}
\begin{equation}
  \betappn - 1 =\frac{1}{2} \frac{\alpha^2 \beta}{{(1+\alpha^2)}^2}\,,
  \qquad \gammappn - 1= -2\frac{\alpha^2}{1+\alpha^2}\,,
\end{equation}
where $\alpha$ and $\beta$ are defined by
\begin{equation}
\begin{aligned}
  \alpha & = \sqrt{2} \dfrac{\partial \ln A}{\partial \chibar} \simeq
  -\dfrac{\sqrt{2 \xi}}{4} \dfrac{e^{2\sqrt{\xi} \chibar}}{1+
      \dfrac{1}{4} e^{2\sqrt{\xi} \chibar}}\,, \\ 
\beta & =
2 \frac{\partial^2 \ln A}{\partial \chibar^2} \simeq - \xi
\dfrac{e^{2\sqrt{\xi} \chibar}}{\left(1+\dfrac{1}{4} e^{2\sqrt{\xi}
    \chibar} \right)^2}\,.
\end{aligned}
\end{equation}
Using the limit $\xi \phibar^2 = \Mpbar^2/M^2 \gg 1$ for $\chibar$
\begin{equation}
e^{\sqrt{\xi} \chibar} = 2 \dfrac{\Mpbar}{M}\, ,
\end{equation}
one gets
\begin{equation}
\begin{aligned}
  \betappn & = 1 + \order{\xi^2 \dfrac{M^2}{\Mpbar^2}}, \\
  \gammappn &= 1-\frac{4\xi}{1+2\xi} + \order{\xi \dfrac{M}{\Mpbar}}.
\end{aligned}
\end{equation}
Thus, $\gammappn$ becomes slightly smaller than unity. The most
stringent bound on $\gammappn$ comes from the Shapiro time delay
measurement using the Cassini spacecraft~\cite{Bertotti:2003rm}:
$-0.03 < (\gammappn-1) \times 10^5 < 4.4$.  This limit translates into
an upper limit on $\xi$ as
\begin{equation}
  \xi < 7.5 \times 10^{-8}.
  \label{eq:shapiro}
\end{equation}

As mentioned in the previous section, the gravitational coupling as
measured by Cavendish-like experiments is $\Mp^2=1/(8\pi G)$, where
$G$ is the measured Newton's constant. It is slightly different from
$\Mpbar^2$ due to the fifth force induced by $\phi$ and reads
\begin{equation}
\Mp^2 = \dfrac{M^2}{A^2 (1+ \alpha^2)} = \dfrac{\Mpbar^2}{1+ 2 \xi} +
\order{M^2} \simeq \Mpbar^2.
\end{equation}
For the values of $\xi$ compatible with the Cassini constraints of
Eq.~\eqref{eq:shapiro}, $\Mp^2$ is therefore indistinguishable from
$\Mpbar^2$ and both quantities will be identified in the following.

Another effect comes from demanding that the potential energy of the
field sources the current acceleration of the Universe. From
Eq.~\eqref{eq:lambda} and $\xi \vev{\phi^2} = \Mpbar^2 \simeq \Mp^2$,
one gets
\begin{equation}
\mphi^2 \simeq 6 \xi \Ho^2 \OmegaL.
\label{eq:mphiDE}
\end{equation}
Therefore, the mass is not a free parameter and for values of $\xi$
satisfying the Cassini bound we get $\mphi < 7 \times 10^{-4} \Ho$,
i.e., the field is extremely light. Let us notice that, because it is
not coupled to other sectors, such a tiny mass is \emph{a priori} not
problematic. Moreover, dynamical mechanisms able to generate small
masses have been proposed; see, for instance,
Ref.~\cite{Nomura:1999py}. The ultra-light scalar field is thus
compatible with all limits associated with an evolution of the
equation of state of dark energy and its
perturbations~\cite{Marsh:2010wq, Hlozek:2014lca}.

\begin{figure}
  \begin{center}
   \includegraphics[width=\figw]{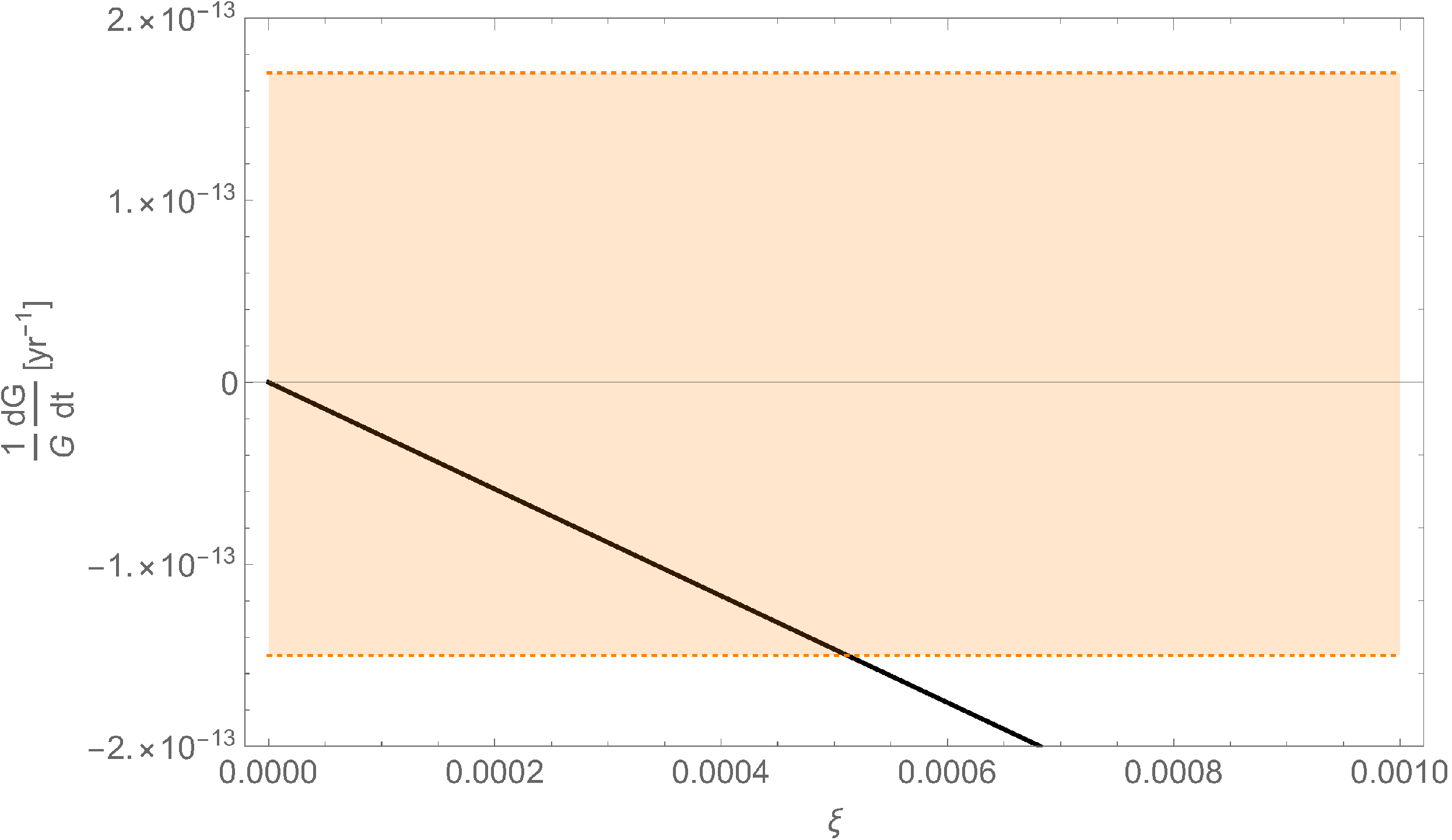}
  \end{center}
  \caption{Time variation of $G$ at present day as a function of
    $\xi$. The bare mass of the field has been set to
    $\mphi=\sqrt{6\xi \OmegaL}\Ho$, the value explaining dark energy
    today. The orange region, which is obtained from improvements in
    the ephemeris of Mars~\cite{Konopliv:2011}, is the observationally
    allowed region.}
  \label{G-variation}
\end{figure}

Finally, there are constraints coming from the cosmological time
variation of $\phi$ which also drives the time variation of the
gravitational constant. The equation of motion of $\phi$ on the
cosmological background is given by
\begin{equation}
 \ddot{\phi} +3 H\dot{\phi} + \mphi^2 \phi -6 \xi \left( 2 H^2+
 \dot{H} \right) \phi=0,
\label{eq:jfevol}
\end{equation}
where a dot stands for a derivative with respect to the cosmic time
and where $\mphi^2$ is given by Eq.~\eqref{eq:mphiDE}. Non-detections
of the time variation of $G$ imply that $\phi$ has not moved
significantly from the initial value until the present epoch. In the
slow-roll regime, the Hubble parameter is approximately given by that
of the standard $\Lambda$CDM model~\cite{2018arXiv180706205P}. Using
this Hubble parameter, we can solve the above equation of motion and
derive the relative time variation of $G$ at present day for different
values of $\xi$. At leading order in $M/\Mpbar$, we have
\begin{equation}
 \dfrac{\dot{G}}{G} = -2 \dfrac{\dot{\phi}}{\phi}\,.
\end{equation}
The result is shown as a thick line in Fig.~\ref{G-variation}.
Interestingly, contrary to the minimally coupled case, the non-minimal
coupling term makes $\phi$ grow, which explains the negative sign of
${\dot G}$.  The orange region is the observationally allowed region
obtained by the improvements in the ephemeris of
Mars~\cite{Konopliv:2011}.  From this figure, we obtain the upper
bound $\xi < 5 \times 10^{-4}$, which is weaker than the one coming
from the Shapiro effect in Eq.~\eqref{eq:shapiro}.

\section{Discussion}
\label{sec:disc}

In the previous sections, we have seen that the ultra-light scalar
field can dynamically generate the large measured value of the Planck
mass $\Mp$ from a much lower gravitational energy scale $M$, which
could be as low as or even smaller than the electroweak scale. Once
its VEV generates the observed Planck mass, the same field can also
source dark energy from its small, but non-vanishing mass
term. However, the mechanism requires a very long period of inflation,
of the order of $\order{\xi^{-1}}$ e-folds. For $\xi < 10^{-7}$, this
means that the scale factor $a$ should have grown during inflation by
a factor of at least the tetration $^{4}e$. Accurate
observations of the cosmic microwave background (CMB) anisotropies in
the last decade strongly support the idea that inflation occurred in
the very early Universe~\cite{Martin:2016oyk, Akrami:2018odb,
  Chowdhury:2019otk}. Although only the last $\sim 60$ e-folds of
inflation can be probed observationally, it is legitimate to suppose
that the total period of inflation that the Universe has experienced
may be much longer. This can happen if the inflaton had a nearly
flat potential over a sufficiently large field range and started its
motion far from the end point of inflation, as this could very well be
the case for the plateau inflationary models favured by the
data~\cite{Martin:2013nzq}.  Another possibility is that the
observable inflation was preceded by a false vacuum phase of the same
field as the one relevant to the last $\sim 60$ e-folds of inflation.
It is also equally possible that the very long inflation is sourced by
a different field than the inflaton responsible for the observable
inflation.

In the following, we describe in more detail the primordial
inflationary part of the model in the presence of the two fields. The
dynamics is easier to understand in the Einstein frame. The equations
of motion for the inflaton field $\psi$ and the canonically normalized
gravity field $\chi$ read~\cite{Ringeval:2005yn}
\begin{equation}
\begin{aligned}
  \dfrac{\deriv{\psibar}{NN} + \sqrt{2} \alpha(\chi) \, \deriv{\chibar}{N}
  \deriv{\psibar}{N}}{3 - \epsilon_1} + \deriv{\psibar}{N} & =
-\dfrac{1}{A^2(\chi)} \dfrac{\ud \ln U}{\ud \psibar}\,,\\
\dfrac{\deriv{\chibar}{NN} -  \left[\alpha(\chi)/\sqrt{2}\right] A^2(\chi) \,
  \deriv{\psibar}{N}^2}{3 - \epsilon_1} + \deriv{\chibar}{N} & =
-\dfrac{\ud \ln U}{\ud \chibar}\,,
\label{eq:evolEF}
\end{aligned}
\end{equation}
where $\epsilon_1$ is the first Hubble flow function in the Einstein
frame
\begin{equation}
\epsilon_1 \equiv -\dfrac{\ud \ln H}{\ud N} = \dfrac{1}{2}
\deriv{\chibar}{N}^2 + \dfrac{1}{2} A^2(\chi) \deriv{\psibar}{N}^2.
\label{eq:eps1EF}
\end{equation}
We have introduced the two-field potential $U(\chi,\psi)$ as
\begin{equation}
U(\chi,\psi) \equiv \dfrac{W(\chi)}{M^2} + \dfrac{A^4(\chi) V(\psi)}{M^2}\,.
\end{equation}
These equations can be simplified by taking the limits we are
interested in, $\xi \ll 1$ and $\xi \phibar^2 \gg 1$ together with
Eq.~\eqref{eq:appfieldEF}. One gets
\begin{equation}
\begin{aligned}
  A^2(\chi) & \simeq 4 e^{-2 \sqrt{\xi} \chibar}, \qquad \alpha(\chi)
  \simeq - \sqrt{2 \xi}\, ,\\
  U(\chi,\psi) & \simeq 2
  e^{-2\sqrt{\xi}\chibar} \left[\dfrac{\mphi^2}{\xi} + \dfrac{8
      V(\psi)}{M^2} e^{-2\sqrt{\xi} \chibar} \right].
\end{aligned}
\label{eq:potapprox}
\end{equation}
From these equations, with $\mphi^2 / \xi =\order{\Ho^2}$, one gets
\begin{equation}
U(\chi,\psi) \simeq  16 \, e^{-4 \sqrt{\xi} \chibar} \dfrac{V(\psi)}{M^2}\,,
\end{equation}
which from Eq~\eqref{eq:evolEF} gives
\begin{align}
\label{eq:evolpsiEF}
  \dfrac{\deriv{\psibar}{NN} - 2\sqrt{\xi} \, \deriv{\chibar}{N}
  \deriv{\psibar}{N}}{3 - \epsilon_1} + \deriv{\psibar}{N} & \simeq
-\dfrac{e^{2\sqrt{\xi} \chibar}}{4} \dfrac{\ud \ln V}{\ud \psibar}\,,\\
\label{eq:evolchiEF}
\dfrac{\deriv{\chibar}{NN} + 4\sqrt{\xi} \, e^{-2\sqrt{\xi} \chibar}
  \deriv{\psibar}{N}^2}{3 - \epsilon_1} + \deriv{\chibar}{N} & \simeq 4 \sqrt{\xi}\,.
\end{align}
Under the slow-roll approximation, one can find an approximate
solution of Eqs.~\eqref{eq:evolpsiEF} and
\eqref{eq:evolchiEF}. Let us first assume in
Eq.~\eqref{eq:evolchiEF} that
\begin{equation}
4 \sqrt{\xi} \, e^{-2\sqrt{\xi} \chibar} \deriv{\psibar}{N}^2 \ll
\deriv{\chibar}{N}.
\label{eq:hypsmall}
\end{equation}
The slow-roll solution for $\chibar$ reads
\begin{equation}
\deriv{\chibar}{N} \simeq 4 \sqrt{\xi} \ll 1.
\label{eq:chisol}
\end{equation}
This equation implies that $\chibar(N) \propto 4\sqrt{\xi} N$. As can
be explicitly checked by using Eq.~\eqref{eq:appfieldEF}, this is the
Einstein frame manifestation of the tachyonic growth of
$\phi$. Plugging the above equation into Eq.~\eqref{eq:evolpsiEF}, we
get the slow-roll solution for the inflaton $\psi$ (with $\xi \ll 1$)
\begin{equation}
\deriv{\psibar}{N} \simeq -\dfrac{e^{2\sqrt{\xi} \chibar}}{4}
\dfrac{\ud \ln V}{\ud \psibar} \,.
\end{equation}
This allows us to estimate the first Hubble flow function from
Eq.~\eqref{eq:eps1EF}
\begin{equation}
\epsilon_1 \simeq 8 \xi + \dfrac{e^{2\sqrt{\xi} \chibar}}{8}
\left(\dfrac{\ud \ln V}{\ud \psibar} \right)^2.
\label{eq:epsone}
\end{equation}
Under our hypothesis~\eqref{eq:hypsmall}, the second term
\begin{equation}
\epsilon_{\psi} \equiv \dfrac{e^{2\sqrt{\xi} \chibar}}{8}
\left(\dfrac{\ud \ln V}{\ud \psibar} \right)^2,
\label{eq:epspsi}
\end{equation}
is small, and for $\xi \ll 1$, we recover the condition of slow-roll
inflation $\epsilon_1 \ll 1$. Let us mention that reversing the
inequality in our working hypothesis of Eq.~\eqref{eq:hypsmall} is not
acceptable as one would get a value larger than unity for $\epsilon_1$
and no inflation at all.

From Eq.~\eqref{eq:epsone}, we see that the tachyonic growth of $\phi$
induces corrections to the inflaton dynamics, compared to what one would
have obtained in standard GR. The factor $e^{2\sqrt{\xi} {\bar \chi}}$
in Eq.~\eqref{eq:epspsi} increases with $\chi$ and this implies that the
term $\epsilon_{\psi}$ will ultimately dominate in
Eq.~\eqref{eq:epsone}. When this happens, the kinetic energy of the
$\psi$-field will drive inflation towards its graceful ending, as
needed. Let us notice that, even if the the first Hubble flow function
$\epsilon_1$ has an additional term, $8\xi$, a more detailed
calculation shows that the tensor-to-scalar ratio is given by $r=16
\epsilon_{\psi*}$, which passes current constraints for plateau-like
potentials~\cite{Array:2015xqh, Akrami:2018odb}.

A last comment is in order concerning the very large-scale structure
of the Universe generated in this scenario. Although not explicit in
the above description, the fact that the inflaton potential $V(\psi)$
should be asymptotically very flat implies that not only $\phi$ but
also $\psi$ is expected to develop large super-Hubble fluctuations. In
that situation, the earliest phase of inflation is certainly chaotic,
and possibly eternal, depending on the shape of
$V(\psi)$~\cite{Vilenkin:1983xq, Linde:1983mx, Linde:1986fd,
  1986PhLB..175..395L, Goncharov:1987ir}. Determining the probability
that the chaotic regime ends in a classical evolution matching our
scenario is still an open and relevant question, which we leave to
future work~\cite{GarciaBellido:1994vz, Vennin:2016wnk}.

\section{Conclusion}
\label{sec:conc}

We have proposed a novel scenario where both the Planck scale and dark
energy are dynamically generated by the stochastic and tachyonic
motion of a weakly non-minimally coupled ultra-light scalar field,
which alleviates the large hierarchy between the Planck, electroweak
scale, neutrino mass scale, and cosmological constant scales.
According to this scenario, such an ultra-light field is still present
in the current Universe and mediates a long-range fifth force among
bodies.  Cosmological observations and Solar-System experiments
require $\xi$ to be small.  The stronger bound comes from the Shapiro
effect measured by the Cassini spacecraft and $\xi <
\order{10^{-7}}$. Generically, all improvements on the bounds of a
possible non-minimal coupling in terrestrial or Solar-System
environment will be relevant in constraining, or proving, our
model~\cite{Will:2014kxa}.

However, we could think of other means to test the scenario. A
possible route of detection could be through the cosmological motion
of the scalar field, which is not exactly static. The
equation-of-state parameter $w$ for dark energy differs from $-1$ due
to the slow motion of the field as
\begin{equation}
w=-1+\order{\xi}.
\end{equation}
According to Ref.~\cite{Sprenger:2018tdb}, one could expect the future
Euclid satellite~\cite{2013LRR....16....6A} and SKA radio
telescope~\cite{2015aska.confE..12P}, combined with Planck CMB data,
to constrain the deviation of $w+1$ down to $10^{-3}$. This will
certainly not be enough to reach the current bound $\xi <
\order{10^{-7}}$ and one may have to wait for the next generation of
giant radio telescopes~\cite{Tegmark:2008au}. However, let us remark
that as soon as the field $\phi$ starts to evolve on cosmological
scales, the effective gravitational coupling given by $\xi \phi^2$ is
also modified. We have not assessed the possible joint constraints from
varying dark energy and a varying Newton's constant, but it may be
another interesting route to explore.

Recent detections of gravitational waves (GWs) by the LIGO/VIRGO
observatory~\cite{Abbott:2016blz} have opened a new era for GW
astronomy.  In the future, various types of GW detectors will be
launched and the physics of the gravity sector will be probed much
more widely and deeply.  It has been shown in Ref.~\cite{Yagi:2009zz}
that it is possible to place an upper limit on the Brans-Dicke
parameter $\omegaBD \gtrsim 4\times 10^8$ using the Deci-hertz
Interferometer Gravitational wave Observatory (DECIGO), which is a
planned space-based GW detector consisting of four constellations of
three satellites forming a triangular shape~\cite{Seto:2001qf}.  In
the massless limit, the non-minimal coupling parameter is related to
$\omegaBD$ as $\xi = 1/(4\omegaBD)$.  From the DECIGO limit, we obtain
$\xi \lesssim 6\times 10^{-10}$, which is a roughly $2$
orders-of-magnitude improvement over the current bound.  Hence, there
is a window that can be probed by future GW experiments such as DECIGO
or LISA~\cite{Will:2004xi}.

\begin{acknowledgments}
The work of C.R is supported by the ``Fonds de la Recherche Scientifique
- FNRS'' under Grant $\mathrm{N^{\circ}T}.0198.19$.  This work is
supported by JSPS Grant-in-Aid for Young Scientists (B) No.15K17632
(T.S.), by the MEXT Grant-in-Aid for Scientific Research on
Innovative Areas No.15H05888 (T.S., M.Y.), No.17H06359 (T.S.),
No.18H04338 (T.S.), and No.18H04579 (M.Y.), by the JSPS KAKENHI Grant
Numbers JP25287054 (M.Y.) and JP18K18764 (M.Y.), and by the Mitsubishi
Foundation (M.Y.).
\end{acknowledgments}

\bibliographystyle{apsrev}
\bibliography{biblio}

\end{document}